\documentclass[times,11pt]{article}
\usepackage{amsmath}
\usepackage{amsthm}
\usepackage{amsfonts}
\usepackage{latexsym}
\usepackage{amssymb}
\usepackage{xspace}

\newtheorem{theorem}{Theorem}
\newtheorem{claim}[theorem]{Claim}
\newtheorem{lemma}[theorem]{Lemma}
\newtheorem{defn}{Definition}
\newtheorem{cor}[theorem]{Corollary}

\newcommand\E{\mbox{\bf E}}

\def\reals{\mathbb{R}}
\def\ball{\mathbb{B}_n}
\def\simplex{\mathbb{S}_n}
\def\bS{\mathbb{S}}
\def\cube{\mathbb{C}_n}

\def\mP{\mathcal{P}}
\def\mF{\mathcal{F}}
\def\mA{\mathcal{A}}

\newcommand{\ignore}[1]{}
\newcommand{\equaltri}{\triangleq}
\def\poracle{{\sc Separation Oracle}\xspace}
\def\doracle{{\sc Optimization Oracle}\xspace}

\def\oalg{{\sc OnlineAlg}\xspace}
\def\popt{{\sc PrimalGameOpt}\xspace}
\def\dopt{{\sc DualGameOpt}\xspace}
\def\pdopt{{\sc PrimalDualGameOpt}\xspace}

\def\ons{{\sc Online Newton Step}\xspace}
\def\sproj{{\sc SimplexProject}\xspace}

\newcommand\mycases[4] {{
\left\{
\begin{array}{ll}
    {#1} & {#2} \\\\
    {#3} & {#4}
\end{array}
\right. }}
\newcommand\mythreecases[6] {{
\left\{
\begin{array}{ll}
    {#1} & {#2} \\\\
    {#3} & {#4} \\\\
    {#5} & {#6}
\end{array}
\right. }}

\title{Approximate Convex Optimization by Online Game Playing}
\date{}
\author{Elad Hazan \thanks{Part of this research was supported by Sanjeev Arora's NSF grants MSPA-MCS 0528414, CCF 0514993, ITR
0205594} \\
IBM Almaden Research Center\\
hazan@us.ibm.com}

\begin{document}

\maketitle

\begin{abstract}

Lagrangian relaxation and approximate optimization algorithms have
received much attention in the last two decades. Typically, the
running time of these methods to obtain a $\varepsilon$
approximate solution is proportional to $\frac{1}{\varepsilon^2}$.
Recently, Bienstock and Iyengar, following Nesterov, gave an
algorithm for fractional packing linear programs which runs in
$\frac{1}{\varepsilon}$ iterations. The latter algorithm requires
to solve a convex quadratic program every iteration - an
optimization subroutine which dominates the theoretical running
time.

We give an algorithm for convex programs with strictly convex
constraints which runs in time proportional to
$\frac{1}{\varepsilon}$. The algorithm does {\bf not} require to
solve any quadratic program, but uses gradient steps and
elementary operations only. Problems which have strictly convex
constraints include maximum entropy frequency estimation,
portfolio optimization with loss risk constraints, and various
computational problems in signal processing.

As a side product, we also obtain a simpler version of Bienstock
and Iyengar's result for general linear programming, with similar
running time.

We derive these algorithms using a new framework for deriving
convex optimization algorithms from online game playing
algorithms, which may be of independent interest.

\end{abstract}

\section{Introduction}

The design of efficient approximation algorithms for certain convex and linear
programs has received much attention in the previous two decades. Since
interior point methods and other polynomial time algorithm are often too slow
in practice \cite{bienstock01potential}, researchers have tried to design
approximation algorithms. Shahrokhi and Matula \cite{matula} developed the
first approximation algorithm for the maximum concurrent flow problem. Their
result spurred a great deal of research, which generalized the techniques to
broader classes of problems (linear programming, semi-definite programming,
packing and covering convex programs) and improved the running time
\cite{leighton,KPST,PST,grigor-first,GK,Fleischer,grigoriadis-kachiyan,KleinLu,AHKsdp}.

In this paper we consider approximations to more general convex
programs. The convex feasibility problem we consider is of the
following form (the optimization version can be reduced to this
feasibility problem by binary search),
\begin{align}
f_j (x) & \leq 0  \quad \forall j \in [m] \label{eqn:general} \\
x & \in \simplex \notag
\end{align}

Where $\{f_j , j \in [m]\}$ is a (possibly infinite) set of convex
constraints and $\simplex = \{x \in \reals^n , \sum_i x_i = 1 ,
x_i \geq 0\}$ is the unit simplex. Our algorithm work almost
without change if the simplex is replaced by other simple convex
bodies such as the ball or hypercube. The more general version,
where $\simplex$ is replaced by an arbitrary convex set in
Euclidian space, can also be handled at the expense of slower
running time (see section \ref{subsection:scp}).

We say that an algorithm gives an $\varepsilon$-approximate
solution to the above program if it returns $x \in \mP$ such that
$\forall j \in [m] \ . \ f_j(x) \leq \varepsilon$, or returns
proof that the program is infeasible. Hence, in this paper we
consider an additive notion of approximation. A multiplicative
$\varepsilon$-approximation is a $x \in \mP$ such that $\forall j
\in [m] \ . \ f_j(x) \leq \lambda^* ( 1 + \varepsilon)$ where
$\lambda^* = \min_{x \in \mP} \max_{i \in [m]} f_i(x)  $. There
are standard reductions which convert an additive approximation
into a multiplicative approximation. Both of these reductions are
orthogonal to our results and can be applied to our algorithms.
The first is based on simple scaling, and is standard in previous
work (see \cite{PST,YoungRounding,MWsurvey}) and increases the
running time by a factor of $\frac{1}{\lambda^*}$. For the special
case fractional packing and covering problems, there is a
different reduction based on binary search which increases the
running time only by a poly-logarithmic factor
\cite{bienstock,nesterov}.

A common feature to all of the prior algorithms is that they can
be viewed, sometimes implicitly, as Frank-Wolfe \cite{FrankWolfe}
algorithms, in that they iterate by solving an optimization
problems over $\simplex$ (more generally over the underlying
convex set), and take convex combinations of iterates. The
optimization problem that is iteratively solved is of the
following form.

\begin{eqnarray*}
\forall p \in \bS_m \ . \ \mbox{\doracle(p)}  \equaltri &
\mycases{x \in \simplex {\mbox { s.t }} \sum_j p_j f_j(x) \leq 0}
{\mbox{ if exists such $x$}} {FAIL} {{\mbox{ otherwise}}}
\end{eqnarray*}

It is possible to extend the methods of PST \cite{PST} and others
to problems such as (\ref{eqn:general}) (see
\cite{jansen,khandekarPhD}) and obtain the following theorem.
Henceforth $\omega$ stands for the {\it width} of the instance
--- a measure of the size of the instance numbers --- defined as
$\omega = \max_{j \in [m]} \max_{x \in \simplex} f_i(x) - \min_{j
\in [m]} \min_{x \in \simplex} f_i(x)$.

\begin{theorem}[previous work] \label{thm:previouswork1}
There exists an algorithm that for any $\varepsilon
> 0$, returns a $\varepsilon$-approximation solution to
mathematical program (\ref{eqn:general}). The algorithm makes at
most $ \tilde{O}(\frac{\omega^2}{\varepsilon^2 } )$ calls to
\doracle, and requires $O(m)$ time between successive oracle
calls.
\end{theorem}

{\bf Remark 1:} Much previous work focuses on reducing the
dependance of the running time on the width. Linear dependence on
$\omega$ was achieved for special cases such as {\it packing and
covering problems} (see \cite{YoungRounding}). For covering and
packing problems the dependence on the width can be removed
completely, albeit introducing another $n$ factor into the running
time \cite{jansen}. These results are orthogonal to ours, and it
is possible that the ideas can be combined.

{\bf Remark 2:} In case the constraint functions are linear,
\doracle can be implemented in time $O(mn)$. Otherwise, the oracle
reduces to optimization of a convex non-linear function over a
convex set.

\medskip

Klein and Young \cite{KleinYoung} proved an $\Omega(\varepsilon^{-2})$ lower
bound for Frank-Wolfe type algorithms for covering and packing linear programs
under appropriate conditions. This bound applies to all prior lagrangian
relaxation algorithms till the recent result of Bienstock and Iyengar
\cite{bienstock}. They give an algorithm for solving packing and covering
linear programs in time linear in $\frac{1}{\varepsilon}$, proving

\begin{theorem}[\cite{bienstock}]
There exists an algorithm that for any $\varepsilon
> 0$, returns a $\varepsilon$-approximation solution to packing or
covering linear programs with $m$ constraints. The algorithm makes
at most $ \tilde{O}(\frac{{n}}{\varepsilon })$ iterations. Each
iteration requires solving a convex separable quadratic program.
The algorithm requires $O(mn)$ time between successive oracle
calls.
\end{theorem}

Their algorithm has a non-combinatorial component, viz., solving
convex separable quadratic programs. To solve these convex
programs one can use interior point methods, which have large
polynomial running time largely dominating the entire running time
of the algorithm. The \cite{bienstock} algorithm is based on
previous algorithms by Nesterov \cite{nesterov} for special cases
of linear and conic programming. Nesterov's algorithm pre-computes
a quadratic program, which also dominates the running time of his
algorithm.

\subsection{Our results}

We give a simple approximation algorithms for convex programs
whose running time is linear in $\frac{1}{\varepsilon}$. The
algorithms requires only gradient computations and combinatorial
operations (or a separation oracle more generally), and does not
need to solve quadratic programs.

The $\Omega(\varepsilon^{-2})$ lower bound of Klein and Young is circumvented
by using the strict convexity of the constraints. The constraint functions are
said to be strictly convex if there exists a positive real number $H > 0$ such
that $\min_{j \in [m]} \min_{x \in \mP} \nabla^2 f_j(x) \succeq H \cdot I $
\footnote{we denote $A \succeq B$ if the matrix $A- B \succeq 0$ is positive
semi-definite} . In other words, the Hessian of the constraint function is
positive definite (as opposed to positive semi-definite) with smallest
eigenvalue at least $H
> 0$.

Our running time bounds depend on the gradients of the constraint
functions as well. Let $G = \max_{j \in [m]} \max_{x \in \mP}
\|\nabla f_j(x) \|_2$ be an upper bound on the norm of the
gradients of the constraint functions. $G$ is related to the width
of the convex program: for linear constraints, the gradients are
simply the coefficients of the constraints, and the width is the
largest coefficient. Hence, $G$ is at most $\sqrt{n}$ times the
width. In section (\ref{section:applications}) we prove the
following Theorem.

\begin{theorem}[Main 1] \label{thm:scp}
There exists an algorithm that for any $\varepsilon > 0$, returns
a $\varepsilon$-approximate solution to mathematical program
(\ref{eqn:general}). The algorithm makes at most $
\tilde{O}(\frac{G^2}{H } \cdot \frac{1}{\varepsilon} )$ calls to
\poracle, and requires a single gradient computation and
additional $O(n)$ time between successive oracle calls.
\end{theorem}

{\bf Remark:} Commonly the gradient of a given function can be
computed in time which is linear in the function representation.
Examples of functions which admit linear-time gradient computation
include polynomials, logarithmic functions and exponentials.

\smallskip

The separation oracle which our algorithm invokes is defined as
\begin{eqnarray*}
\forall x \in \simplex \ . \ \mbox{\poracle(x)}  \equaltri &
\mycases{  \ {\mbox {$j \in [m]$ s.t }} f_j(x) > \varepsilon}
{\mbox{ if exists such $f_j$}} {FAIL} {{\mbox{ otherwise}}}
\end{eqnarray*}

If the constraints are given explicitly, often this oracle is easy to implement
in time linear in the input size. Such constraints include linear functions,
polynomials and logarithms. This oracle is also easy to implement in parallel:
the constraints can be distributed amongst the available processors and
evaluated in parallel.

For all cases in which $H$ is zero or too small the theorem above
cannot be applied. However, we can apply a simple reduction to
strictly convex constraints and obtain the following corollary.

\begin{cor} \label{cor:scp}
For any $\varepsilon > 0$, there exists an algorithm that returns
a $\varepsilon$-approximate solution to mathematical program
(\ref{eqn:general}). The algorithm makes at most $
\tilde{O}(\frac{G^2}{\varepsilon^2 } )$ calls to \poracle and
requires additional $O(n)$ time and a single gradient computation
between successive oracle calls.
\end{cor}

In comparison to Theorem~\ref{thm:previouswork1}, this corollary
may require $O(n)$ more iterations. However, each iteration
requires a call to \poracle, as opposed to \doracle. A \poracle
requires only function evaluation, which can many times be
implemented in linear time in the input size, whereas an \doracle
could require expensive operations such as matrix inversions.

There is yet another alternative to deal with linear constraints and yet obtain
linear dependence on $\varepsilon$. This is given by the following theorem. The
approximation algorithm runs in time linear in $\frac{1}{\varepsilon}$, and yet
does not require a lower bound on $H$. The downside of this algorithm is the
computation of ``generalized projections". A generalized projection of a vector
$y \in \reals^n$ onto a convex set $\mP$ with respect to PSD matrix $A \succeq
0$ is defined to be $\prod^A _\mP(y) = \arg\min_{x \in \mP} (x - y)^\top A
(x-y)$. Generalized projections can be cast as convex mathematical programs. If
the underlying set is simple, such as the ball or simplex, then the program
reduces to a convex quadratic program.
\begin{theorem}[Main 2] \label{thm:lp}
There exists an algorithm that for any $\varepsilon > 0$ returns a
$\varepsilon$-approximate solution to mathematical program
\eqref{eqn:general}. The algorithm makes at most $
\tilde{O}(\frac{n G }{\varepsilon } )$ calls to \poracle and
requires computation of a generalized projection onto $\simplex$,
a single gradient computation and additional $\tilde{O}(n^2)$ time
between successive oracle calls.
\end{theorem}

An example of an application of the above theorem is the following
linear program.
\begin{equation}
\forall j \in [m] \ . \ A_j \cdot x  \ge  0, \quad x \in \simplex
\label{lp1}
\end{equation}
It is shown in \cite{DV} that general linear programming can be reduced to this
form, and that without loss of generality, $\forall j \in [m] \ \ \|A_j\|=1$.
This format is called the ``perceptron" format for linear programs. As a
corollary to Theorem \ref{thm:lp}, we obtain

\begin{cor} \label{cor:lp}
There exists an algorithm  that for any $\varepsilon > 0$
returns a $\varepsilon$-approximate solution to linear program
(\ref{lp1}). The algorithm makes $ \tilde{O}(\frac{n}{\varepsilon}
)$ iterations. Each  iteration requires $\tilde{O}(n( m+n))$
computing time plus computation of a generalized projection onto
the simplex.
\end{cor}

Theorem \ref{thm:lp} and Corollary \ref{cor:lp} extend the result
of Bienstock and Iyengar \cite{bienstock} to general convex
programming~\footnote{Bienstock and Iyengar's techniques can also
be extended to full linear programming by introducing dependence
on the width which is similar to that of our algorithms
\cite{biens}.}. The running time of the algorithm is very similar
to theirs: the number of iterations is the same, and each
iteration also requires to solve convex quadratic programs
(generalized projections onto the simplex in our case). Our
algorithm is very different from \cite{bienstock}. The analysis is
simpler, and relies on recent results from online learning. We
note that the algorithm of Bienstock and Iyengar allows improved
running time for sparse instances, whereas our algorithm currently
does not.

\subsection{Lagrangian relaxation and solving zero sum games} \label{subsection:online}

The relation between lagrangian relaxation and solving zero sum games was
implicit in the original PST work, and explicit in the work of Freund and
Schapire on online game playing \cite{FSgames} (the general connection between
zero sum games and linear programming goes back to von Neumann).

Most previous lagrangian relaxation algorithms can be viewed as
reducing the optimization problem at hand to a zero sum game, and
then applying a certain online game playing algorithm, the
Multiplicative Weights algorithm, to solve the game.

Our main insight is that the Multiplicative Weights algorithm can
be replaced by any {\it online convex optimization} (see next
section for precise definition) algorithm. Recent developments in
online game playing introduce algorithms with much better
performance guarantees for online games with convex payoff
functions \cite{AH,HKKA}. Our results are derived by reducing
convex optimization problems to games with payoffs which stem from
convex functions, and using the new algorithms to solve these
games.

The online framework also provides an alternative explanation to
the aforementioned Klein and Young $\Omega(\varepsilon^{-2})$
lower bound on the number of iterations required by Frank-Wolfe
algorithms to produce an $\varepsilon$-approximate solution.
Translated to the online framework, previous algorithm were based
on online algorithms with $\Omega(\sqrt{T})$ {\it regret} (the
standard performance measure for online algorithms, see next
section for precise definition). Our linear dependance on
$\frac{1}{\varepsilon}$ is the consequence of using of online
algorithms with $O(\log T)$ regret. This is formalized in Appendix
\ref{section:lowerbounds}.

\section{The general scheme}

We outline a general scheme for approximately solving convex
programs using online convex optimization algorithms. This is a
generalization of previous methods which also allows us to derive
our results stated in the previous section.

For this section we consider the following general mathematical
program, which generalizes (\ref{eqn:general}) by allowing an
arbitrary convex set $\mP$.
\begin{align}
f_j (x) & \leq 0  \quad \forall j \in [m] \label{eqn:general2} \\
x & \in \mP \notag
\end{align}

In order to approximately solve (\ref{eqn:general2}), we reduce the
mathematical problem to a game between two players: a {\it primal player} who
tries to find a feasible point and the {\it dual player} who tries to disprove
feasibility. This reduction is formalized in the following definition.

\begin{defn}
The associated game with mathematical program (\ref{eqn:general2})
is between a primal player that plays $x \in \mP$ and a dual
player which plays a distribution over the constraints $p \in
S_m$. For a point played by the primal player and a distribution
of the dual player, the loss that the primal player incurs (and
the payoff gained by the dual player) is given by the following
function
\begin{equation}
\forall \ x \in \mP \ ,  p \in \bS_m \ . \  \ g(x,p) \equaltri
\sum_j p_j f_j(x) \notag
\end{equation}
The value of this game is defined to be $ \lambda^*
\equaltri \min_{x \in \mP} \max_{p \in \bS_m} g(x,p) $.
Mathematical program (\ref{eqn:general2}) is feasible iff
$\lambda^* \leq 0$.
\end{defn}

By the above reduction, in order to check feasibility of
mathematical program (\ref{eqn:general2}), it suffices to compute
the value of the associated game $\lambda^*$. Notice that the game
loss/payoff function $g$ is smooth over the convex sets $S_m$ and
$\mP$, linear with respect to $p$ and convex with respect to $x$.
For such functions, generalizations to the von Neumann minimax
theorem, such as \cite{sion} \footnote{All algorithms and theorems
in this paper can be proved without relying on this minimax
theorem. In fact, our results provide a new algorithmic proof of
the generalized min-max theorem which is included in Appendix
\ref{sec:minimax}.} imply that
$$\lambda^* = \min_{x \in \mP} \max_{p \in \bS_m} g(x,p) = \max_{p \in \bS_m} \min_{x \in \mP}
g(x,p)$$

This suggests a natural approach to evaluate $\lambda^*$: simulate
a repeated game between the primal and dual players such that  in
each iteration the game loss/payoff is determined according to the
function $g$. In the simulation, the players play according to an
online algorithm.

The online algorithms we consider fall into the {\it online convex
optimization} framework \cite{zinkevich}, in which there is a
fixed convex compact feasible set $\mP \subset \reals^n$ and an
{\em arbitrary, unknown} sequence of convex cost functions
$f_1,f_2,\ldots:\mP \rightarrow \reals$. The decision maker must
make a sequence of decisions, where the $t^\text{th}$ decision is
a selection of a point $x_t \in \mP$ and there is a cost of
$f_t(x_t)$ on period $t$.  However, $x_t$ is chosen with only the
knowledge of the set $\mP$, previous points $x_1,\ldots,x_{t-1}$,
and the previous functions $f_1,\ldots,f_{t-1}$. The standard
performance measure for online convex optimization algorithms is
called {\it regret} which is defined as:
\begin{eqnarray} \label{eqn:regret}
\mbox{Regret}(\mA,T)  \equaltri  \sup_{f_1,...,f_T} \left\{
\sum_{t=1}^T f_t(x_t) - \min_{x^* \in \mP} \sum_{t=1}^T f_t(x^*)
\right\}
\end{eqnarray}
We say that an algorithm $\mA$ has {\it low regret} if
$\mbox{Regret}(\mA,T) = o(T)$. Later, we use to the procedure
\oalg, by which we refer to any low regret algorithm for this
setting.

Another crucial property of online convex optimization algorithms
is their running time. The running time is the time it takes to
produce the point $x_t \in \mP$ given all prior game history.

The running time of our approximate optimization algorithms will
depend on these two parameters of online game playing algorithms:
regret and running time. In Appendix \ref{sec:online} we survey
some of the known online convex optimization algorithms and their
properties.

\medskip

We suggest three methods for approximating (\ref{eqn:general2}) using the
approach outlined above. The first ``meta algorithm" (it allows freedom in
choice for the implementation of the online algorithm) is called \popt and
depicted in figure \ref{fig:meta}. For this approach, the dual player is
simulated by an optimal adversary: at iteration $t$ it plays a dual strategy
$p_t$ that achieves at least the game value $\lambda^*$ (this reduces exactly
to \poracle).

The implementation of the primal player is an online convex optimization
algorithm with low regret, which we denote by \oalg. This online convex
optimization algorithm produces decisions which are points in the convex set
$\mP$. The cost functions $f_1,f_2,\ldots:\mP \rightarrow \reals$ are
determined by the dual player's distributions. At iteration $t$, if the
distribution output by the dual player us $p_t$, then the cost function to the
online player is
$$ \forall x \in \mP \ . \ f_t(x) \equaltri g(x,p_t)$$

The low-regret property of the online algorithm used ensures that in the long
run, the average strategy of the primal player will converge to the optimal
strategy. Hence the average loss will converge to $\lambda^*$.

The ``dual" version of this approach, in which the dual player is
simulated by an online algorithm and the primal by an oracle, is
called \dopt. In this case, the adversarial implementation of the
primal player reduces to \doracle. The dual player now plays
according to an online algorithm \oalg. This online algorithm
produces points in the $m$-dimensional simplex - the set of all
distributions over the constraints. The payoff functions are
determined according to the decisions of the primal player: at
iteration $t$, if primal player produced point $x_t \in \mP$, the
payoff function is
$$ \forall p \in \bS_m \ . \ f_t(p) \equaltri g(x_t,p)$$

We also explore a third option, in which both players are
implemented by online algorithms. This is called the \pdopt
meta-algorithm. Pseudo-code for all versions is given in figure
(\ref{fig:meta}).

\begin{figure}[h!]
\hrule\hrule\vspace{2pt}  {\bf \popt($\varepsilon$)}

Let $t \leftarrow 1$. While Regret(\oalg,t) $ \geq \varepsilon t$
do
\begin{itemize}
\item
Let $x_t \leftarrow $ \oalg($p_1,...,p_{t-1}$).
\item
Let $j \leftarrow$ \poracle($x_t$). If $FAIL$ return $x_t$. Let
$p_t \leftarrow e_j$, where $e_j$ is the $j$'th standard basis
vector of $\reals^n$.
\item
$t \leftarrow t + 1$
\end{itemize}
Return $\bar{p} = \frac{1}{T} \sum_{t=1}^T p_t$\\

\hrule\hrule\vspace{2pt}  {\bf \dopt($\varepsilon$)}

Let $t \leftarrow 1$. While Regret(\oalg,t) $ \geq \varepsilon t$
do
\begin{itemize}
\item
Let $p_t \leftarrow $ \oalg($x_1,...,x_{t-1}$).
\item
Let $x_t \leftarrow$ \doracle($p_t$). If $FAIL$ return $p_t$.
\item
$t \leftarrow t + 1$
\end{itemize}
Return $\bar{x} \equaltri \frac{1}{T} \sum_{t=1}^T x_t$\\

\hrule\hrule\vspace{2pt}  {\bf \pdopt($\varepsilon$)}

Let $t \leftarrow 1$. While Regret(\oalg,t) $ \geq
\frac{\varepsilon}{2} t$ do
\begin{itemize}
\item
Let $x_t \leftarrow $ \oalg($p_1,...,p_{t-1}$).
\item
Let $p_t \leftarrow $ \oalg($x_1,...,x_{t-1}$).
\item
$t \leftarrow t + 1$
\end{itemize}
If $\bar{x} \equaltri \frac{1}{T} \sum_{t=1}^T x_t$ is
$\varepsilon$-approximate return $\bar{x}$. Else, return $\bar{p} =
\frac{1}{T} \sum_{t=1}^T p_t$. \\

\hrule
 \caption{meta algorithms for approximate optimization by online game playing \label{fig:meta}}
\end{figure}

The following theorem shows that all these approaches yield an
$\varepsilon$-approximate solution when the online convex
optimization algorithm used to implement \oalg has low regret.

\begin{theorem} \label{thm:general}
Suppose $\oalg$ is an online convex optimization algorithm with low regret.
If a solution to mathematical program (\ref{eqn:general2}) exists, then
meta-algorithms \popt, \dopt and \pdopt return an $\varepsilon$-approximate
solution. Otherwise, \popt and \dopt return a dual solution proving that the
mathematical program is infeasible, and \pdopt returns a dual solution proving
the mathematical program to be $\varepsilon$-close to being infeasible.

Further, a $\varepsilon$-approximate solution is returned in
$O(\frac{R}{\varepsilon})$ iterations, where $R = R(\oalg,\varepsilon)$ is the
smallest number $T$ which satisfies the inequality $\mbox{Regret}(\oalg,T) \leq
\varepsilon T$.
\end{theorem}

\begin{proof}

{\bf Part 1: correctness of \popt} \\

If at iteration $t$ \poracle returns $FAIL$, then by definition of
\poracle,
$$ \forall p^* \ . \ g(x_t,p^*) \leq \varepsilon \ \Rightarrow \ \forall j \in [m] \ . \ f_j({x}_t) \leq \varepsilon$$
implying that $x_t$ is a $\varepsilon$-approximate solution.

Otherwise, for every iteration $g(x_t,p_t) > \varepsilon$, and we
can construct a dual solution as follows. Since the online
algorithm guarantees sub-linear regret, for some iteration $T$ the
regret will be $R \leq \varepsilon T$. By definition of regret we
have for any strategy $x^* \in \mP$,

\begin{eqnarray*} \label{eqn:thmpart1}
\varepsilon < \frac{1}{T} \sum_{t=1}^T g(x_t,p_t) \leq \frac{1}{T}
\sum_{t=1}^T g(x^*,p_t) + \frac{R}{T} \leq \frac{1}{T}
\sum_{t=1}^T g(x^*,p_t) + \varepsilon \leq g(x^*,\bar{p}) +
\varepsilon
\end{eqnarray*}

Where the last inequality is by the concavity (linearity) of
$g(x,p)$ with respect to $p$ Thus,
$$ \forall x^* \ . \ g(x^*,\bar{p}) > 0  $$

Hence $\bar{p}$ is a dual solution proving that the mathematical
program is infeasible.

\medskip

{\bf Part 2: correctness of \dopt} The proof of this part is
analogous to the first, and given in the full version of this
paper.

If for some iteration $t$ \doracle returns $FAIL$. According to
the definition of \doracle,
$$ \forall x \in \mP \ . \ g(x,p_t) > 0 $$
implying that $p_t$ is a dual solution proving the mathematical
program to be infeasible.

Else, in every iteration $g(x_t,p_t) \leq 0$. As before, for some
iteration $T$ the regret of the online algorithm will be $R \leq
\varepsilon T$. By definition of regret we have (note that this
time the online player wants to maximize his payoff)

$$ \forall p^* \in P(\mF) \ . \ 0 \geq \frac{1}{T} \sum_{t=1}^T g(x_t,p_t) \geq \frac{1}{T} \sum_{t=1}^T g(x_t,p^*)
- \frac{R}{T} \geq \frac{1}{T} \sum_{t=1}^T g(x_t,p^*)
-\varepsilon
$$

Changing sides and using the convexity of the function $g(x,p)$
with respect to $x$ (which follows from the convexity of the
functions $f \in \mF$) we obtain (for $\bar{x} = \frac{1}{T}
\sum_{t=1}^T x_t$)
$$ \forall p^* \in P(\mF) \ . \  g( \bar{x} , p^*) \leq \frac{1}{T} \sum_{t=1}^T g(x_t,p^*) \leq \varepsilon $$
Which in turn implies that
$$  \forall f \in \mF \ . \ f(\bar{x}) \leq  \varepsilon $$

Hence $\bar{x}$ is a $\varepsilon$-approximate solution.

\medskip

{\bf Part 3: correctness of \pdopt}

Denote $R_1,R_2$ the regrets attained by both online algorithms
respectively. Using the low regret properties of the online
algorithms we obtain for any $x^*,p^*$

\begin{eqnarray} \label{eqn:thmpart3}
 \forall x^*,p^* \ . \  \sum_{t=1}^T g(x_t,p^*) - R_1 \leq
\sum_{t=1}^T g(x_t,p_t) \leq  \sum_{t=1}^T g(x^*,p_t) + {R_2}
\end{eqnarray}
Let $x^*$ be such that $\forall p \in P(\mF) \ . \  g(x^*,p) \leq
\lambda^*$. By convexity of $g(x,p)$ with respect to $x$,
$$  \forall p^* \ . \ g(\bar{x},p^*) \leq \frac{1}{T} \sum_{t=1}^T g(x_t,p^*) \leq \frac{1}{T} \sum_{t=1}^T g(x^*,p_t) +  \frac{R_2 + R_1}{T} \le \lambda^* +  {\varepsilon}  $$
Similarly, let $p^*$ be such that $\forall x \in \mP \ . \
g(x,p^*) \geq \lambda^*$. Then by concavity of $g$ with respect to
$p$ and equation \ref{eqn:thmpart3} we have
$$  \forall x^* \ . \ g(x^*,\bar{p}) \geq \frac{1}{T} \sum_{t=1}^T g(x^*,p_t) \geq \frac{1}{T} \sum_{t=1}^T g(x_t,p^*)  - \frac{R_2 + R_1}{T} \geq \lambda^* - {\varepsilon} $$

Hence, if $\lambda^* \leq 0$, then $\bar{x}$ satisfies
$$ \forall p^* \ . \ g(\bar{x},p^*) \leq \varepsilon \ \Rightarrow \ \forall j \in [m] \ . \ f_j(\bar{x}) \leq \varepsilon$$

And hence is a $\varepsilon$-approximate solution. Else,
$$ \forall x^* \ . \ g({x^*},\bar{p}) > - \varepsilon $$
And $\bar{p}$ is a dual solution proving that the following
mathematical program is infeasible.
\begin{align*}
f_j (x) & \leq -\varepsilon  \quad \forall j \in [m]  \\
x & \in \mP \notag
\end{align*}
\end{proof}

\section{Applications} \label{section:applications}

\subsection{Strictly convex programs} \label{subsection:scp}

We start with the easiest and perhaps most surprising application
of Theorem \ref{thm:general}. Recall that the feasibility problem
we are considering:
\begin{align} \label{eqn:formulation:scp}
 f_j (x) & \leq 0  \quad \forall j \in [m] \\
x & \in \simplex \notag
\end{align}
Where the functions $\{f_j\}$ are all strictly convex such that
$\forall x \in \simplex , j \in [m] \ . \  \nabla^2 f_j(x) \succeq
H \cdot I_n$ and $ \| \nabla f_j(x)  \|_2 \leq G$

\begin{figure}[h!]
\hrule\hrule\vspace{2pt}  {\bf {\sc StrictlyCovexOpt}.} \\
Input: Instance in format \eqref{eqn:formulation:scp}, parameters
$G,H$ approximation guarantee $\varepsilon$.
\\
Let $t \leftarrow 1 \ , \ x_1 \leftarrow \frac{1}{n}\vec{1}$. \\
While $t \leq \frac{G^2}{H} \frac{1}{\varepsilon}\log
\frac{1}{\varepsilon}$ do
\begin{itemize}
\item
Let $j \leftarrow$ \poracle($x_t$) (i.e. an index of a violated
constraint). If all constraints are satisfied return $x_t$. Else,
let $\nabla_{t-1} = \nabla f_j(x_{t-1})$. Let $p_t \leftarrow e_j$
where $e_j$ is the $j$'th standard basis vector of $\reals^n$.

\item
Set $y_t = x_{t-1} - \frac{1}{H \cdot t} \nabla_{t-1} $

\item
Set $x_t = \ $\sproj$(y_t)$.

\item
$t \leftarrow t + 1$
\end{itemize}
Return $\bar{p} = \frac{1}{T} \sum_{t=1}^T p_t$\\

\hrule \caption{An approximation algorithm for strictly convex
programs. Here $\vec{1}$ stands for the vector with one in all
coordinates.  \label{figure:scp}}
\end{figure}

\begin{proof}[Proof of Theorem \ref{thm:scp}]

Consider the associated game with value
\begin{equation*}
\lambda^* \equaltri \min_{x \in \simplex} \max_{j \in [m]} f_j(x)
\end{equation*}

The convex problem is feasible iff $\lambda^* \leq 0$. To
approximate $\lambda^*$, we apply the \popt meta algorithm. In
this case, the vectors $x_t$ are points in the simplex, and $p_t$
are distributions over the constraints. The online algorithm used
to implement \oalg is Online Convex Gradient Descent (OCGD). The
resulting algorithm is strikingly simple, as depicted in figure
\ref{figure:scp}.

According to Theorem 1 in \cite{HKKA}, the regret of OCGD is
bounded by $Regret(T) = O(\frac{G^2 }{H} \log T)$. Hence, the
number of iterations till the regret drops to $\varepsilon T$ is
$\tilde{O}(\frac{G^2}{H} \frac{1}{\varepsilon})$. According to
Theorem \ref{thm:general}, this is the number of iterations
required to obtain an $\varepsilon$-approximation.

In each iteration, the OCGD algorithm needs to update the current
online strategy (the vector $x_t$) according to the gradient and
project onto $\simplex$. This requires a single gradient
computation. A projection of a vector $y \in \reals^n$ onto
$\simplex$ is defined to be $\prod _\mP(y) = \arg\min_{x \in
\simplex} \|x - y\|_2$. The projection of a vector onto the
simplex can be computed in time $\tilde{O}(n)$ (see procedure
\sproj described in Appendix \ref{subsection:projections}). Other
than the gradient computation and projection, the running time of
OCGD is $O(n)$ per iteration.
\end{proof}
{\bf Remark:} It is clear that the above algorithm can be applied
the more general version of convex program (\ref{eqn:general2}),
where the simplex is replaced by an arbitrary convex set $\mP
\subseteq \reals^n$. The only change required is in the projection
step. For Theorem \ref{thm:scp}, we assumed the underlying convex
set is the simplex, hence the projection can be computed in time
$\tilde{O}(n)$. Projections can be computed in linear time also
for the hypercube and ball. For convex sets which are
intersections of hyperplanes (or convex parabloids), computing a
projection reduces to optimizing a convex quadratic function over
linear (quadratic) constraints. These optimization problems allow
for more efficient algorithms than general convex optimization
\cite{boyd}.

\medskip

As a concrete example of the application of Theorem \ref{thm:scp}, consider the
case of strictly convex quadratic programming. In this case, there are $m$
constraint functions of the form $f_j(x) = x^\top A_j x + b_j^\top x + c$,
where the matrices $A_j$ are positive-definite. If $A_j \succeq H \cdot I$, and
$\forall _{x \in \simplex} \|A_j x + b_j\|_2 \leq G$, then Theorem
\ref{thm:scp} implies that an $\varepsilon$-approximate solution can be found
in $\tilde{O}(\frac{G^2 }{H \varepsilon})$ iterations.

The implementation of \poracle involves finding a constraint
violated by more than $\varepsilon$. In the worst case all
constrains need be evaluated in time $O(mn^2)$. The gradient of
any constraint can be computed in time $O(n^2)$. Overall, the time
per \poracle computation is $\tilde{O}(mn^2)$. We conclude that
the total running time to obtain a $\varepsilon$-approximation
solution is $\tilde{O}(\frac{G^2  m n^2}{H \varepsilon})$. Notice
that the input size is $m n^2$ in this case.

\subsection{Linear and Convex Programs} \label{sec:lp}

In this section we prove Theorem \ref{thm:lp}, which gives an
algorithm for convex programming that has running time
proportional to $\frac{1}{\varepsilon}$. As a simple consequence
we obtain corollary \ref{cor:lp} for linear programs. The
algorithm is derived using the \popt meta-algorithm and the \ons
(ONS) online convex optimization algorithm (see appendix
\ref{sec:online}) to implement \oalg. The resulting algorithm is
described in figure \ref{figure:convexalg} below.

\begin{figure}[h!]
\hrule\hrule\vspace{2pt}  {\bf {\sc CovexOpt}.} \\
Input: Instance in format \eqref{eqn:concaveprogram}, parameters
$G,D,\omega$ approximation guarantee $\varepsilon$.
\\
Let $t \leftarrow 1 \ , \ x_1 \leftarrow \frac{1}{n}\vec{1} \ ,
\beta \leftarrow \frac{1}{2}\min\{1,\frac{1}{4GD}\} \ , \  A_0 \leftarrow \frac{1}{D^2 \beta^2} I_n \ , \ A_0^{-1} \leftarrow {D^2 \beta^2} I_n $. \\
While $t \leq {6 n G D } \frac{1}{\varepsilon}\log
\frac{1}{\varepsilon}$ do

\begin{itemize}
\item
Let $j \leftarrow$ \poracle($x_t$) (i.e. an index of a violated
constraint). If all constraints are satisfied return $x_t$. Else,
let $\nabla_{t-1} = \nabla \{\log(e + \omega^{-1} f_j (x))\}$ and
$p_t \leftarrow e_j$ where $e_j$ is the $j$'th standard basis
vector of $\reals^n$.

\item
Set $y_t = x_{t-1} + \frac{1}{\beta} A_{t-1}^{-1} \nabla_{t-1} $

\item
Set $x_t = \mathop{\arg\min}_{x \in \mP}\ (y_t-x)^\top
A_{t-1}(y_t-x) $

\item

Set  $A_{t} = A_{t-1} + \nabla_{t-1} \nabla_{t-1}^\top$, and
$A_t^{-1} = A_{t-1}^{-1} - \frac{A_{t-1}^{-1} \nabla_{t-1}
\nabla_{t-1}^\top A_{t-1} ^{-1}}{1 + \nabla_{t-1}^\top
A_{t-1}^{-1} \nabla_{t-1}}$
\end{itemize}
Return $\bar{p} = \frac{1}{T} \sum_{t=1}^T p_t$\\

\hrule
 \caption{An approximation algorithm for convex
programs. Here $I_n$ stands for the $n$-dimensional identity
matrix. \label{figure:convexalg}}
\end{figure}

Since for general convex programs the constraints are not strictly
convex, one cannot apply online algorithms with logarithmic regret
directly as in the previous subsection. Instead, we first perform
a reduction to a mathematical program with exp-concave
constraints, and then approximate the reduced instance.

\begin{proof}[Proof of Theorem \ref{thm:lp}]
In this proof it is easier for us to consider concave constraints
rather than convex. Mathematical program \eqref{eqn:general} can
be converted to the following by negating each constraint:
\begin{align}
 f_j (x) & \geq 0  \quad \forall j \in [m] \label{eqn:concaveprogram} \\
x & \in \mP \notag
\end{align}
where the functions $\{f_j\}$ are all concave such that $\forall x
\in \mP , j \in [m] \ . \   \| \nabla f_j(x)  \|_2 \leq G$ and
$\forall x \in \mP, j \in [m] \ . \   | f_j(x) | \leq \omega$.
This program is even more general than \eqref{eqn:general} as it
allows for an arbitrary convex set $\mP$ rather than $\simplex$.

Let $\rho = \max_{x \in \mP} \min_j \{f_j(x)\} $. The question to
whether this convex program is feasible is equivalent to whether
$\rho > 0$.

In order to approximately solve this convex program, we consider a
different concave mathematical program,
\begin{align}
\log(e + \omega^{-1} f_j (x)) & \geq 1  \quad \forall j \in [m] \label{eqn:logprogram} \\
x & \in \mP \notag
\end{align}
It is a standard fact that concavity is preserved for the
composition of a non-decreasing concave function with another
concave function, i.e. the logarithm of positive concave functions
is itself concave. To solve this program we consider the
(non-linear) zero sum game defined by the following min-max
formulation

\begin{equation}
\lambda^* \equaltri \max_{x \in \mP} \min_{j \in [m]} \log(e +
\omega^{-1} f_j(x) )  \label{mp3}
\end{equation}

The following two claims show that program \eqref{eqn:logprogram}
is closely related to \eqref{eqn:concaveprogram}.

\begin{claim}
$\lambda^* = \log(e + \omega^{-1} \rho)$.
\end{claim}
\begin{proof}
Let $x$ be a solution to \eqref{eqn:concaveprogram} which achieves
the value $\rho$, that is $\forall j \in [m] \ . \ f_j(x) \geq
\rho$. This implies that $\forall j \in [m] \ . \ \log(e +
\omega^{-1} f_j(x)) \geq \log(e + \omega^{-1} \rho) $, and in
particular $\forall q \ \ g(x,q) \geq \log(e+\omega^{-1}\rho)$
hence $\lambda^* \geq \log(e + \omega^{-1}\rho)$.

For the other direction, suppose that $\lambda^* = \log(e + z) >
\log(e+ \omega^{-1}\rho) $ for some $z > \omega^{-1}\rho$. Then
there exists an $x$ such that $\forall j \in [m] \ . \ \log(e
+\omega^{-1} f_j(x)) \geq \lambda^* > \log(e + z)$ or equivalently
$\forall j \in [m] \ . \ f_j(x) \geq z > \rho$ in contradiction to
the definition of $\rho$.
\end{proof}

\begin{claim} \label{claim:lpofflineapprox}
An $\varepsilon$-approximate solution for \eqref{eqn:logprogram}
is a $3 \omega \varepsilon$-approximate solution for
\eqref{eqn:concaveprogram}.
\end{claim}
\begin{proof}
A $\varepsilon$-approximate solution to \eqref{eqn:logprogram}
satisfies $\forall j \ . \ \log(e +  \omega^{-1} f_j(x)) \geq
\lambda^* - \varepsilon = \log(e+\omega^{-1}\rho) - \varepsilon $.
Therefore, by monotonicity of the logarithm we have
\begin{eqnarray*}
\omega^{-1} f_j(x) & \geq  e^{\log(e+\omega^{-1}\rho) - \varepsilon} - e \\
& =  (e + \omega^{-1}\rho) \cdot e^{-\varepsilon} - e \\
&\geq  (e + \omega^{-1}\rho) (1 - \varepsilon)  - e & \mbox{ since $e^{-x} \geq 1 - x$ }\\
& = \omega^{-1} \rho(1 - \varepsilon) - e \varepsilon
\end{eqnarray*}
Which implies
$$  f_j(x) \geq  \rho(1 -  \varepsilon) - 3 \omega \varepsilon$$

\end{proof}

We proceed to approximate $\lambda^*$ using \popt and choose the
\ons (ONS) algorithm (see appendix \ref{sec:online}) as \oalg. The
resulting algorithm is depicted in figure \ref{figure:convexalg}.

We note that here the primal player is maximizing payoff as
opposed to the minimization version in the proof of Theorem
\ref{thm:general}. The maximization version of Theorem
\ref{thm:general} can be proved analogously.

In order to analyze the number of iterations required, we
calculate some parameters of the constraints of formulation
\eqref{eqn:logprogram}. See appendix \ref{sec:online} for
explanation on how the different parameters effect the regret and
running time of \ons.

The constraint functions are $1$-exp-concave, since their
exponents are linear functions. Their gradients are bounded by
$$ \tilde{G}  \equaltri \max_{j \in m} \max_{x \in \mP} \|\nabla \log(e + \omega^{-1} f_j(x))\| =
\max_{j \in m} \max_{x \in \mP} \|\frac{ \omega^{-1} \nabla
f_j(x)}{e + \omega^{-1} f_j(x)}\| \leq \omega^{-1} G
$$

According to Theorem 2 in \cite{HKKA}, the regret of ONS is
$O((\frac{1}{\alpha} + {G}D) n \log T)$. In our setting, $\alpha =
1$ and $G$ is replaced by $\tilde{G}$. Therefore, the regret
becomes smaller then $\varepsilon T$ after $O(\frac{n G D
\omega^{-1} }{\varepsilon})$ iterations. By
Theorem~\ref{thm:general}, after $T = \tilde{O}(\frac{n G D
\omega^{-1} }{\delta})$ iterations we obtain an
$\delta$-approximate solution, i.e a solution $x^*$ such that
$$ \min_{j \in [m]} \log(e + \omega^{-1} f_j(x^*)) \geq \lambda^* - \delta$$
Which by claim \ref{claim:lpofflineapprox} is a $3 \omega
\delta$-approximate solution to the original math program. Taking
$\delta= O(\omega^{-1} \varepsilon)$ we obtain an
$\varepsilon$-approximate solution to concave program
\eqref{eqn:concaveprogram} in $T = \tilde{O}(\frac{n G
D}{\varepsilon})$ iterations.

We now analyze the running time per iteration. Each iteration
requires a call to \poracle in order to find an
$\varepsilon$-violated constraint. The gradient of the constraint
need be computed.  According to the gradient the ONS algorithm
takes $O(n^2)$ time to update its internal data structures (which
are $y_t,A_t, A_t^{-1}$ in figure \ref{figure:convexalg}). Finally
ONS computes a generalized projection onto $\mP$, which
corresponds to computing $\mathop{\arg\min}_{x \in \mP}\
(y-x)^\top A_{t-1}(y-x) $ given $y$ (see appendix
\ref{subsection:projections})

If $\mP = \simplex$, then $D = 1$ and the bounds of Theorem
\ref{thm:lp} are met.

\end{proof}

Given Theorem \ref{thm:lp}, it is straightforward to derive
corollary \ref{cor:lp} for linear programs:

\begin{proof}[proof of Corollary \ref{cor:lp}]
For linear programs in format \eqref{lp1}, the gradients of the
constraints are bounded by $\max_{j \in [m]} \|A_j\| \leq 1$. In
addition, \poracle is easy to implement in time $O(mn)$ by
evaluating all constraints.

Denote by $T^S_{proj}$ the time to compute a generalized
projection onto the simplex. A worst case bound is $T^S_{proj} =
O(n^3)$, using interior point methods (this is an instance
quadratically constrained convex quadratic program, see
\cite{boyd}).

Plugging these parameters into Theorem \ref{thm:lp}, the total
running time comes to

$$ \tilde{O}(\frac{n }{\varepsilon} \cdot (nm + n^2 + T^S_{proj}) )$$

\end{proof}

{\bf Remark:} As is the case for strictly convex programming, our
framework actually provides a more general algorithm that requires
a \poracle. Given such an oracle, the corresponding optimization
problem can be solved in time $ \tilde{O}(\frac{n}{\varepsilon}
\cdot (n^2 + T_{A,proj} + T_{oracle}))$ where $T_{oracle}$ is the
running time of \poracle.

\subsection{Derivation of previous results}

For completeness, we prove Theorem \ref{thm:previouswork1} using
our framework. Even more generally, we prove the  theorem for
general convex program \eqref{eqn:general2} rather than
\eqref{eqn:general}.

\begin{proof}[Proof of Theorem \ref{thm:previouswork1}]
Consider the associated game with value
\begin{equation*}
\lambda^* \equaltri \min_{x \in \mP} \max_{j \in [m]} f_j(x) =
\max_{p \in S_m} \min_{x \in \mP} \sum_{i=1}^m p_i f_i(x)
\end{equation*}
The convex problem is feasible iff $\lambda^* \leq 0$. To
approximate $\lambda^*$, we apply the \dopt meta algorithm. The
vectors $x_t$ are points in the convex set $\mP$, and $p_t$ are
distributions over the constraints, i.e. points in the $m$
dimensional simplex. The payoff functions for \oalg in iteration
$t$ are of the form
$$ \lambda p \ . \  g(x_t,p) = \sum_i p_i f_i(x_t) $$

The online algorithm used to implement \oalg is the Multiplicative
Weights algorithm (MW). According to Theorem \ref{thm:mwregret} in
appendix \ref{sec:online}, the regret of MW is bounded by
$\mbox{Regret}_T(MW) = O({G_\infty }\sqrt{T \log m})$ (the
dimension of the online player is $m$ in this case). Hence, the
number of iterations till the regret drops to $\varepsilon T$ is
$\tilde{O}(\frac{G_\infty^2}{\varepsilon^2})$. According to
Theorem \ref{thm:general}, this is the number of iterations
required to obtain an $\varepsilon$-approximation.

To bound $G_\infty$, note that the payoff functions $\lambda p \ .
\  g(x_t,p) $ are linear. Their gradients are $m$-dimensional
vectors such that the $i$'th coordinate is the value of the $i$'th
constraint on the point $x_t$, i.e.  $f_i(x_t)$. Thus, the
$\ell_\infty$ norm of the gradients can be bounded by
$$ G_\infty = \max_{x \in \mP} \max_{t \in [T]} \nabla (\lambda p \ . \  g(x_t,p))  \leq \max_{i \in [m]} \max_{x \in \mP} f_i(x) $$
And the latter expression is bounded by the width $\omega =
\max_{i \in [m]} \max_{x \in \mP} |f_i(x)| $. Thus the number of
iterations to obtain an $\varepsilon$-approximate solution is
bounded by $\tilde{O}(\frac{\omega^2}{\varepsilon^2})$.

In each iteration, the MW algorithm needs to update the current
online strategy (the vector $p_t$) according to the gradient in
time $O(m)$. This requires a single gradient computation.
\end{proof}

\section{Acknowledgements}

Many thanks to Sanjeev Arora for numerous helpful suggestions. We
would also like to thank Noga Alon, Satyen Kale and Nimrod Megiddo
for helpful discussions.

\bibliographystyle{alpha}
\bibliography{neps}

\newcommand{\etalchar}[1]{$^{#1}$}
\begin{thebibliography}{HKKA06}

\bibitem[AH05]{AH}
Amit Agarwal and Elad Hazan.
\newblock Efficient algorithms for online game playing and universal portfolio
  management.
\newblock {\em ECCC TR06-033}, 2005.

\bibitem[AHK05a]{MWsurvey}
S.~Arora, E.~Hazan, and S.~Kale.
\newblock The multiplicative weights update method: a meta algorithm and
  applications.
\newblock {\em Manuscript}, 2005.

\bibitem[AHK05b]{AHKsdp}
Sanjeev Arora, Elad Hazan, and Satyen Kale.
\newblock Fast algorithms for approximate semide.nite programming using the
  multiplicative weights update method.
\newblock In {\em 46th IEEE FOCS}, pages 339--348, 2005.

\bibitem[Ber06]{bertsimas}
Dimitris Bertsimas.
\newblock personal communications, 2006.

\bibitem[BI04]{bienstock}
D.~Bienstock and G.~Iyengar.
\newblock Solving fractional packing problems in oast(1/\&\#949;) iterations.
\newblock In {\em STOC '04: Proceedings of the thirty-sixth annual ACM
  symposium on Theory of computing}, pages 146--155, New York, NY, USA, 2004.
  ACM Press.

\bibitem[Bie01]{bienstock01potential}
D.~Bienstock.
\newblock Potential function methods for approximately solving linear programs:
  Theory and practice, 2001.

\bibitem[Bie06]{biens}
Daniel Bienstock.
\newblock personal communications, 2006.

\bibitem[Cov91]{cover}
T.~Cover.
\newblock Universal portfolios.
\newblock {\em Math. Finance}, 1:1--19, 1991.

\bibitem[DS06]{miro}
M.~Dudik and R.~E. Schapire.
\newblock Maximum entropy distribution estimation with generalized
  regularization.
\newblock In {\em Proceedings of the 19th Annual Conference on Learning
  Theory}, pages 123--138, 2006.

\bibitem[DV04]{DV}
John Dunagan and Santosh Vempala.
\newblock A simple polynomial-time rescaling algorithm for solving linear
  programs.
\newblock In {\em STOC '04: Proceedings of the thirty-sixth annual ACM
  symposium on Theory of computing}, pages 315--320, New York, NY, USA, 2004.
  ACM Press.

\bibitem[Fle00]{Fleischer}
Lisa~K. Fleischer.
\newblock Approximating fractional multicommodity flow independent of the
  number of commodities.
\newblock {\em SIAM J. Discret. Math.}, 13(4):505--520, 2000.

\bibitem[FS99]{FSgames}
Y.~Freund and R.~E. Schapire.
\newblock Adaptive game playing using multiplicative weights.
\newblock {\em Games and Economic Behavior}, 29:79--103, 1999.

\bibitem[FW56]{FrankWolfe}
M.~Frank and P.~Wolfe.
\newblock An algorithm for quadratic programming.
\newblock {\em Naval Research Logistics Quarterly}, 3:149--154, 1956.

\bibitem[GK94]{grigor-first}
Michael~D. Grigoriadis and Leonid~G. Khachiyan.
\newblock Fast approximation schemes for convex programs with many block and
  coupling constraints.
\newblock {\em SIAM Journal on Optimization}, 4:86--107, 1994.

\bibitem[GK95]{grigoriadis-kachiyan}
M.~Grigoriadis and L.~Khachiyan.
\newblock A sublinear-time randomized approximation algorithm for matrix games.
\newblock In {\em Operations Research Letters}, volume~18, pages 53--58, 1995.

\bibitem[GK98]{GK}
N.~Garg and J.~K{\"o}nemann.
\newblock Faster and simpler algorithms for multicommodity flow and other
  fractional packing problems.
\newblock In {\em Proceedings of the 39th Annual Symposium on Foundations of
  Computer Science({FOCS}-98)}, pages 300--309, Los Alamitos, CA, November8--11
  1998. IEEE Computer Society.

\bibitem[HH06]{HH}
Eran Halperin and Elad Hazan.
\newblock Haplofreq - estimating haplotype frequencies efficiently.
\newblock {\em Journal of Computational Biology}, 13(2):481--500, 2006.

\bibitem[HKKA06]{HKKA}
Elad Hazan, Adam Kalai, Satyen Kale, and Amit Agarwal.
\newblock Logarithmic regret algorithms for online convex optimization.
\newblock {\em to appear in 19'th COLT}, 2006.

\bibitem[HSSW96]{helmbold}
David~P. Helmbold, Robert~E. Schapire, Yoram Singer, and Manfred~K. Warmuth.
\newblock On-line portfolio selection using multiplicative updates.
\newblock In {\em ICML}, pages 243--251, 1996.

\bibitem[Jan06]{jansen}
Klaus Jansen.
\newblock Approximation algorithm for the mixed fractional packing and covering
  problem.
\newblock {\em SIAM J. on Optimization}, 17(2):331--352, 2006.

\bibitem[Kha04]{khandekarPhD}
Rohit Khandekar.
\newblock {\em Lagrangian Relaxation based Algorithms for Convex Programming
  Problems}.
\newblock PhD thesis, Indian Institute of Technology, Delhi, 2004.
\newblock Available at \verb+http://www.cse.iitd.ernet.in/~rohitk+.

\bibitem[KL96]{KleinLu}
Philip Klein and Hsueh-I. Lu.
\newblock Efficient approximation algorithms for semidefinite programs arising
  from {MAX CUT} and {COLORING}.
\newblock In {\em Proceedings of the twenty-eighth annual {ACM} Symposium on
  the Theory of Computing}, pages 338--347, 1996.

\bibitem[KPST94]{KPST}
Philip Klein, Serge Plotkin, Clifford Stein, and Eva Tardos.
\newblock Faster approximation algorithms for the unit capacity concurrent flow
  problem with applications to routing and finding sparse cuts.
\newblock {\em SIAM J. Comput.}, 23(3):466--487, 1994.

\bibitem[KW97]{warmuthGDEG}
Jyrki Kivinen and Manfred~K. Warmuth.
\newblock Exponentiated gradient versus gradient descent for linear predictors.
\newblock {\em Inf. Comput.}, 132(1):1--63, 1997.

\bibitem[KY99]{KleinYoung}
Philip Klein and Neal Young.
\newblock On the number of iterations for {Dantzig-Wolfe} optimization and
  packing-covering approximation algorithms.
\newblock {\em Lecture Notes in Computer Science}, 1610:320--327, 1999.

\bibitem[LSM{\etalchar{+}}91]{leighton}
Tom Leighton, Clifford Stein, Fillia Makedon, \&\#201;va Tardos, Serge Plotkin,
  and Spyros Tragoudas.
\newblock Fast approximation algorithms for multicommodity flow problems.
\newblock In {\em STOC '91: Proceedings of the twenty-third annual ACM
  symposium on Theory of computing}, pages 101--111, New York, NY, USA, 1991.
  ACM Press.

\bibitem[LVBL98]{boyd}
Miguel~Sousa Lobo, Lieven Vandenberghe, Stephen Boyd, and Herve Lebret.
\newblock Applications of second-order cone programming, 1998.

\bibitem[Nes04]{nesterov}
Y.~Nesterov.
\newblock Rounding of convex sets and efficient gradient methods for linear
  programming problems.
\newblock Technical Report~4, CORE discussion paper, 2004.

\bibitem[PST91]{PST}
Serge~A. Plotkin, David~B. Shmoys, and Tardos Tardos.
\newblock Fast approximation algorithm for fractional packing and covering
  problems.
\newblock In {\em Proceedings of the 32nd Annual IEEE Symposium on Foundations
  of Computer Science, FOCS'91 (San Juan, Puerto Rico, October 1-4, 1991)},
  pages 495--504, Los Alamitos-Washington-Brussels-Tokyo, 1991. IEEE Computer
  Society Press.

\bibitem[Sio58]{sion}
Maurice Sion.
\newblock On general minimax theorems.
\newblock {\em Pacific J. Math.}, 8:171--176, 1958.

\bibitem[SM90]{matula}
Farhad Shahrokhi and David~W. Matula.
\newblock The maximum concurrent flow problem.
\newblock {\em J. ACM}, 37(2):318--334, 1990.

\bibitem[You95]{YoungRounding}
Neal~E. Young.
\newblock Randomized rounding without solving the linear program.
\newblock In {\em Proceedings of the Sixth Annual {ACM}-{SIAM} Symposium on
  Discrete Algorithms}, pages 170--178, San Francisco, California, 22--24
  January 1995.

\bibitem[Zin03]{zinkevich}
Martin Zinkevich.
\newblock Online convex programming and generalized infinitesimal gradient
  ascent.
\newblock In {\em Proceedings of the Twentieth International Conference
  (ICML)}, pages 928--936, 2003.

\end{thebibliography}

\appendix

\section{Lower bounds} \label{section:lowerbounds}

The algorithmic scheme described hereby generalizes previous
approaches, which are generally known as {\bf Dantzig-Wolfe-type}
algorithms. These algorithms are characterized by the way the
constraints of mathematical program (\ref{eqn:general}) are
accessed: every iteration only a single \doracle call is allowed.

For the special case in which the constraints are linear, there is
a long line of work leading to tight lower bounds on the number of
iterations required for algorithms within the Dantzig-Wolfe
framework to provide an $\varepsilon$-approximate solution.
Already in 1977, Khachiyan proved an
$\Omega(\frac{1}{\varepsilon})$ lower bound on the number of
iterations to achieve an error of $\varepsilon$. This was
tightened to $\Omega(\frac{1}{\varepsilon^2})$ by Klein and Young
\cite{KleinYoung}, and independently by Freund and Schapire
\cite{FSgames}. Some parameters were tightened in \cite{MWsurvey}.

For the game theoretic framework we consider, it is particularly
simple and intuitive to derive tight lower bounds. These lower
bounds do {\bf not} hold for the more general Dantzig-Wolfe
framework. However, virtually all lagrangian-relaxation-type
algorithms known can be derived from our framework. Thus, for all
these algorithms lower bounds on the running time in terms of
$\varepsilon$ can be derived from the following observation.

In our setting, the number of iterations depends on the regret
achievable by the online game playing algorithm which is deployed.
Tight lower bounds are known on regret achievable by online
algorithms.

\begin{lemma}[folklore]
For linear payoff functions any online convex optimization
algorithm incurs $\Omega(G_\infty  \sqrt{T})$ regret.
\end{lemma}
\begin{proof}
This can be seen by a simple randomized example.
 Consider $\mP=[-1,1]$ and linear functions $f_t(x)=r_t x$, where $r_t
= \pm 1$ are chosen in advance, independently with equal
probability. $\E_{r_t}[f_t(x_t)]=0$ for any $t$ and $x_t$ chosen
online, by independence of $x_t$ and $r_t$. However,
$\E_{r_1,\ldots,r_T}[\min_{x \in K} \sum_1^T f_t(x)] =
\E[-|\sum_1^T r_t|] = -\Omega(\sqrt{T})$. Multiplying $r_t$ by any
constant (which corresponds to $G_\infty$) yields the result.
\end{proof}

The above simple lemma is essentially the reason why it took more
than a decade to break the $\frac{1}{\varepsilon^2}$ running time.
The reason why we obtain algorithms with linear dependance on
$\varepsilon$ is the use of strictly convex constraints (or, in
case the original constraints are linear, apply a reduction to
strictly convex constraints).

\section{A general min-max theorem} \label{sec:minimax}

In this section prove a generalized version of the von Neumann
min-max theorem. The proof is algorithmic in nature, and differs
from previous approaches which were based on fixed point theorems.

Freund and Schapire ~\cite{FSgames} provide an algorithmic proof
of the (standard) min-max theorem, and this proof is an extension
of their ideas to the more general case. The additional generality
is in two parameters: first, we allow more general underlying
convex sets, whereas the standard min-max theorem deals with the
$n$-dimensional simplex $\simplex$. Second, we allow
convex-concave functions as defined below rather than linear
functions. Both generalities stems from the fact that we use
general online convex optimization algorithms as the strategy for
the two players, rather than specific ``expert-type" algorithms
which Freund and Schapire use. Other than this difference, the
proof itself follows \cite{FSgames} almost exactly.

The original minimax theorem can be stated as follows.
\begin{theorem}[von Neumann]
If $X,Y$ are finite dimensional simplices and $f$ is a bilinear
function on $X \times Y$, then $f$ has a saddle point, i.e.
$$ \min_{x \in X} \max_{y \in Y} f(x,y) = \max_{y \in Y} \min_{x \in X}  f(x,y)
$$
\end{theorem}

Here we consider a more general setting, in which the two sets
$X,Y$ can be arbitrary closed, non-empty, bounded and convex sets
in Euclidian space and the function $f$ is convex-concave as
defined by:

\begin{defn}
A function $f$ on $X \times Y$ is convex-concave if for every $y
\in Y$ the function $\forall {x \in X} \ \ f_y(x) \equaltri
f(x,y)$ is convex on $X$ and for every $x \in X$ the function
$\forall {y \in Y} \ \ f_x(y) \equaltri  f(x,y)$ is concave on
$Y$.
\end{defn}

\begin{theorem}
If $X,Y$ are closed non-empty bounded convex sets and $f$ is a
convex-concave function on $X \times Y$, then $f$ has a saddle
point, i.e.
$$  \max_{y \in Y} \min_{x \in X}  f(x,y) = \min_{x \in X} \max_{y \in Y} f(x,y)
$$
\end{theorem}

\begin{proof}
Let $\mu^* \equaltri \max_{y \in Y} \min_{x \in X}  f(x,y)$ and
$\lambda^* \equaltri \min_{x \in X} \max_{y \in Y}  f(x,y)$.
Obviously $\mu^* \leq \lambda^*$ (this is called {\it weak
duality}).

Apply the algorithm \pdopt with any low-regret online convex
optimization algorithm. \footnote{for a low-regret algorithm to
exist, we need $f$ to be convex-concave and the underlying sets
$X,Y$ to be convex, nonempty, closed and bounded.} Then by the
regret guarantees we have for the first algorithm (let $\bar{y} =
\frac{1}{T} \sum_{t=1}^T y_t$)
\begin{eqnarray*}
\frac{1}{T} \sum_{t=1}^T f(x_t,y_t)  \leq & \min_{x \in X}
\frac{1}{T} \sum_{t=1}^T f(x,y_t) + \frac{R_1}{T} \\
\leq & \min_{x \in X} f(x,\bar{y}) + \frac{R_1}{T} & \mbox{ concavity of $f_x$} \\
\leq & \max_{y \in Y} \min_{x \in X} f(x,y) + \frac{R_1}{T} \\
= & \mu^* + \frac{R_1}{T}
\end{eqnarray*}

Similarly for the second online algorithm we have (let $\bar{x} =
\frac{1}{T} \sum_{t=1}^T x_t$)
\begin{eqnarray*}
\frac{1}{T} \sum_{t=1}^T f(x_t,y_t)  \geq & \max_{y \in Y}
\frac{1}{T} \sum_{t=1}^T f(x_t,y) - \frac{R_1}{T} \\
\geq & \min_{x \in X} f(\bar{x},y) + \frac{R_1}{T} & \mbox{ convexity of $f_y$} \\
\geq & \min_{x \in X} \max_{y \in Y}  f(x,y) + \frac{R_1}{T} \\
= & \lambda^* + \frac{R_1}{T}
\end{eqnarray*}

Combining both observations we obtain
$$ \lambda^* - \frac{R_2}{T} \leq \mu^* + \frac{R_1}{T} $$
As $T \mapsto \infty$ we obtain $\mu^* \geq \lambda^*$.

\end{proof}

\section{Online convex optimization algorithms} \label{sec:online}

Figure (\ref{fig:algs}) summarizes several known low regret
algorithms. The running time is the time it takes to produce the point $x_t \in
\mP$ given all prior game history.

\begin{figure}[h!]
\begin{center}
\begin{tabular*}{0.8\textwidth}{@{\extracolsep{\fill}}|lll|} \hline
Algorithm & Regret bound & running time \\
\hline
Online convex gradient descent & $ \frac{G^2_2}{H}\log(T) $ & $O(n + T_{proj})$ \\
Online Newton step & $ (\frac{1}{\alpha} + G_2 D) n \log T$ & $O(n^2 + {T}_{A,proj})$ \\
Exponentially weighted online opt.& $\frac{1}{\alpha} n \log(T)$ & $poly(n)$ \\
Multiplicative Weights & $ G_{\infty} \sqrt{T \log n} $ & $O(n)$ \\
\hline
\end{tabular*}
\end{center}
\caption{\label{fig:algs} Various online convex optimization
algorithms and their performance. $T_{proj}$ is the time to
project a vector $y \in \reals^n$ to $\mP$, i.e. to compute
$\arg\min_{x \in \mP} \|y-x\|_2$. $T_{A,proj}$ is the time to
project a vector $y \in \reals^n$ to $\mP$ using the norm defined
by PSD matrix $A$, i.e. to compute $\arg\min_{x \in \mP}
(y-x)^\top A (y-x)$.}
\end{figure}
The first three algorithms are from \cite{HKKA} and are applicable
to the general online convex optimization framework. The
description and analysis of these algorithms is beyond our scope,
and the reader is referred to the paper.

The last algorithm is based on the ubiquitous Multiplicative Weights Update
method (for more applications of the method see survey \cite{MWsurvey}), and is
provided below. Although it was used many times for various applications (for
very detailed analysis in similar settings see \cite{warmuthGDEG}), this
application to general online convex optimization over the simplex seems to be
new (Freund and Schapire \cite{FSgames} analyze this algorithm exactly,
although for linear payoff functions rather than for general convex functions).

\begin{figure}[h!]
\hrule\hrule\vspace{2pt}  {\bf Multiplicative Weights.} \\Inputs: parameter
$\eta < \frac{1}{2}$.
\begin{itemize}
\item
On period 1, play the uniform distribution $x_1 = \vec{1} \in S_n$. Let
$\forall i \in [n] \ . \ w^1_i = 1 $

\item
On period $t$, update
$$ w^t_i = w^{t-1}_i \cdot (1 + \frac{\eta}{G_\infty}\nabla_{t-1}(i)) $$
where $\nabla_t \equaltri \nabla f_t (x_t)$, and play $x_t$ defined as
\begin{align*}
x_t\ \equaltri &\ \frac{w^t}{\|w^t\|_1}
\end{align*}
\end{itemize}
\hrule
 \caption{The Multiplicative Weights algorithm for online convex optimization over the simplex \label{fig:mwalg}}
\end{figure}

This online algorithm, which is called ``exponentiated gradient"
in the machine learning literature, attains similar performance
guarantees to the ``online gradient descent" algorithm of
Zinkevich \cite{zinkevich}. Despite being less general than
Zinkevich's algorithm (we only give an application to the
$n$-dimensional simplex, whereas online gradient descent can be
applied over any convex set in Euclidian space), it attains
somewhat better performance as given in the following theorem.

\begin{theorem}\label{thm:mwregret}
The Multiplicative Weights algorithm achieves the following guarantee, for all
$T \geq 1$.
$$\mbox{Regret}({MW},T)  = \sum_{t=1}^{T} f_t(x_t) -\min_{x\in S_n}\sum_{t=1}^{T}
f_t(x)\ \leq   O(G_\infty   \sqrt{\log n} \sqrt{T}) $$

\end{theorem}

\begin{proof}

Define $\Phi^t = \sum_i w^t_i$. Since $\frac{1}{G_\infty} \nabla_t(i) \in [0,
1]$,
\begin{align*}
\Phi^{t+1}\ =\ & \sum_i w_i^{t+1}
=  \sum_i w_i^t(1 - \frac{\eta}{G_\infty} \nabla_t(i))\ \\
=  &\ \Phi^t - \frac{\eta \Phi^t}{G_\infty}\sum_i x_t(i) \nabla_t(i) & \mbox{ since $x_t(i) = w_i^t/\Phi^t$} \\
= &\ \Phi^t(1 - \eta x_t \nabla_t / G_\infty)\ \\
\le &\  \Phi^te^{-\eta x_t \nabla_t / G_\infty} & \mbox{ since $1 - x \leq
e^{-x} $ for $|x|\leq 1$}
\end{align*}
After $T$ rounds, we have
\begin{align}
\Phi^T \le \Phi^1 e^{-\eta \sum_t x_t \nabla_t / G_\infty} = n e^{-\eta \sum_t
x_t \nabla_t / G_\infty} \label{eqn:EG}
\end{align}
Also, for every $i \in [n]$, using the following facts which follow immediately
from the convexity of the exponential function
\begin{align*}
(1 - \eta)^x \le (1 - \eta x) & \quad \text{if\ } x \in [0, 1] \\
(1 + \eta)^{-x} \le (1 - \eta x) & \quad \text{if\ } x \in [-1, 0]
\end{align*}
We have
\begin{align*}
\Phi^T & = \sum_t w_i^T \ge  w_i^T \\
& =  \prod_t (1 - \eta \nabla_t(i) / G_\infty)
\\
& \geq   (1 - \eta)^{ \sum_{t > 0} \nabla_t(i) / G_\infty} (1 + \eta)^{
\sum_{t < 0} - \nabla_t(i) / G_\infty}
\end{align*}
where the subscripts $\ge 0$ and $< 0$ refer to the rounds $t$ where
$\nabla_t(i)$ is $\ge 0$ and $< 0$ respectively. So together with
(\ref{eqn:EG})

$$ n e^{-\eta \sum_t x_t \nabla_t / G_\infty} \geq  (1 - \eta)^{ \sum_{t > 0} \nabla_t(i) /
G_\infty} (1 + \eta)^{ \sum_{t < 0} - \nabla_t(i) / G_\infty}$$

Taking logarithms and using $\ln({1 \over 1-\eta}) \le \eta + \eta^2$ and
$\ln(1 + \eta) \ge \eta - \eta^2$ for $\eta \le {1 \over 2}$ we get for all $i
\in [n]$ and $x^* \in S_n$

$$ \sum_t x_t \nabla_t \leq (1 + \eta) \sum_{\ge 0} \nabla_t(i) + (1 - \eta) \sum_{< 0} \nabla_t(i)
+ \frac{G_\infty \log n}{\eta} \leq  \sum_t x^* \nabla_t + \eta \sum_t x^*
|\nabla_t| + \frac{G_\infty \log n}{\eta} $$

Where we denote $|\nabla_t|$ for the vector that has in coordinate $i$ the
value $|\nabla_t(i)|$. Therefore
\begin{eqnarray*}
\sum_t f_t(x_t) - f_t (x^*) \leq  \sum_t \nabla_t (x_t - x^*)
\\
\leq \ \eta \sum_t |\nabla_t| x^*  + \frac{G_\infty \log n}{\eta} \\
\leq \ \eta T G_\infty   + \frac{G_\infty \log n}{\eta} \\
\end{eqnarray*}
And the proof follows choosing $\eta = \sqrt{\frac{{\log n}}{{T}}}$

\end{proof}

Remark: As the algorithm is phrased, it needs to know $T$ and $G_\infty$ in
advance (this is not a problem for the way we use online algorithms as a
building block in approximate optimization). Standard techniques can be used so
that the algorithm need not accept any input: the dependence on $T$ can be
removed by doubling the value of $T$ as it is being exceeded. The dependence on
$G_\infty$ can be removed by using, at any point in the algorithm application,
the largest $G_\infty$ value encountered thus far.

\section{Projections onto convex sets}
\label{subsection:projections}

Many of the algorithms for online convex optimization described in
this chapter require to compute projections onto the underlying
convex set. This correspond to the following computational
problem: given a convex set $\mP \subseteq \reals^n$, and a point
$y \in \reals^n$, find the point in the convex set which is
closest in Euclidian distance to the given vector. We denote the
latter by $\Pi_{\mP}[y]$.

This problem can be formulated as a convex program, and thus
solved in polynomial time by interior point methods or the
ellipsoid method. However, for many simple convex bodies which
arise in practical applications (some of which will be detailed in
following chapters), projections can be computed much more
efficiently. For the $n$-dimensional unit sphere, cube and the
simplex these projections can be computed combinatorially in
$\tilde{O}(n)$ time, rendering the online algorithms much more
efficient when applied to these convex bodies.

\paragraph{The unit sphere}

The simplex projection is over the unit $n$-dimensional sphere,
which we denote by $\ball = \{x \in \reals^n \ , \ \|x\|_2 \leq
1\}$. Given a vector $y \in \reals^n$, it is easy to verify that
it's projection is
$$ \Pi_{\mP}[y] = \mycases{y}{\|y\| \leq
1}{\frac{y}{\|y\|}}{o/w}$$

\paragraph{The unit cube}

Another body which is easy to project onto is the unit
$n$-dimensional cube, which we denote by $\cube^n = \{x \in
\reals^n \ , \ \|x\|_\infty \leq 1\}$ (i.e. each coordinate is
less than or equal to one). Given a vector $y \in \reals^n$, it is
easy to verify that it's projection is
$$ \forall i \in [n] \ . \ \Pi_{\mP}[y](i) = \mythreecases{y[i]}{y(i)
\in [-1,1]}{1}{y(i) > 1}{-1}{y(i) < -1} $$

\paragraph{The Simplex}

The first non-trivial projection we encounter is over the
$n$-dimensional simplex. The simplex is the set of all
$n$-dimensional distributions, and hence is particularly
interesting in many real-world problems, portfolio management and
haplotype frequency estimation just to name a few. Surprisingly,
given an arbitrary vector in Euclidian space, the closest
distribution can be found in near linear time. A procedure for
computing such a projection is given in figure
\ref{figure:simplexproject}.

\begin{figure}[h!]
\hrule\hrule\vspace{2pt}  {\bf \sproj(y).} \\
Suppose w.l.o.g that $y_1 \leq y_2 ... \leq y_n$ (otherwise sort
indices of $y$).
\begin{itemize}
\item Let $a \in \reals$ be the number such that
$ \sum_{i=1}^n \max\{y_i - a,0\} = 1 $. Set\\ $\forall i \in [n] \
. \ x_i = \max\{y_i - a,0\}$.

\item
{\bf Return} x
\end{itemize}

\hrule \caption{A Procedure for projecting onto the Simplex
\label{figure:simplexproject}}
\end{figure}

\begin{lemma}
\sproj$(y)$ is the projection of $y \in \reals^n$ to the
$n$-dimensional simplex, and can be computed in time
$\tilde{O}(n)$.
\end{lemma}
\begin{proof}
First, note that the number $a$ computed in \sproj exists and is
unique. This follows since the function $f(a) = \sum_{i=1}^n
\max\{y_i - a,0\}$ is continuous, monotone decreasing, and takes
values in $[0,\infty)$.

Next, the vector returned $x = \sproj(y)$ is in the simplex. All
its coordinates are positive by definition, and $\sum_{i=1}^n x_i
= \sum_{i=1}^n  \max\{y_i -a,0\} = 1$.

To show that $x$ is indeed the projection we need to prove that it
is the optimum of the mathematical program
$$ \min_{x \in S_n} \sum_{i=1}^n (y_i - x_i)^2 $$

It suffices to show that $x$ is a local optimum, since the program
is convex. Let $c_i \equaltri y_i - x_i$. Then the values
$\{c_i\}$ are decreasing and of the form
$$ (c_1,...,c_{n}) = (a,...,a,y_k,...,y_n)$$

An allowed local change is of the form $x'_{i} \leftarrow x_{i} -
\varepsilon$ and $x'_{j} \leftarrow x_{j} + \varepsilon$ for $i <
j$, since all coordinates larger than $k$ have $x_k = 0$. This
would cause a change in the objective of the form
$$ \sum_{i=1}^d (y_i - x_i)^2  - (y_i - x'_{i})^2 = a^2 - (a + \varepsilon)^2 + c_j - (c_j - \varepsilon)^2 = -2(a - c_j) \varepsilon -2 \varepsilon^2 < 0 $$

Hence would only reduce the objective. Therefore $x$ is indeed the
projection of $y$.

The procedure \sproj requires sorting $n$ elements, and finding
the value $a$, which is standard to implement in $O(n \log n) =
\tilde{O}(n)$ time.

\end{proof}

\section{Examples of strictly convex mathematical programs}

In this section we give some examples of problems which arise in practice and
contain strictly convex constraints. The first example henceforth, and many
others, appear in the excellent survey of \cite{boyd}.

\subsection{Portfolio optimization with loss risk constraints}

A classical portfolio problem described in \cite{boyd} is to
maximize the return of a portfolio over $n$ assets under
constraints which limit its risk. The underlying model assumes a
gaussian distribution of the asset prices with known
$n$-dimensional mean and covariance matrix.

The constraints bound the probability of the portfolio to achieve
a certain return under the model. A feasibility version, of just
checking whether a portfolio exists that attains certain risk with
different mean-covariance parameters, can be written as the
following mathematical program
\begin{align}
p_j^\top x  - \beta  \cdot x^\top \Sigma_j \ x  & \geq \alpha   \quad \forall j \in [m] \label{eqn:portfoliorisk} \\
x & \in \simplex \notag
\end{align}
We refer the reader to \cite{boyd} section 3.4 for more details.

If the underlying gaussian distributions are not degenerate, the
covariance matrices $\Sigma_j$ are positive definite. If the
covariance matrices are degenerate - there is a linear dependance
between two or more assets. In this case it is sufficient to
consider a smaller portfolio with only one of the assets.

The non-degeneracy translates to a strictly positive constant $H
> 0$ such that $\forall j \in [m] \ . \ \Sigma_j \succeq H \cdot
I$. This is, of course, the smallest eigenvalue of the covariance
matrices.

\subsection{Computing the best CRP in hindsight with transaction
costs normalization}

In a popular model for portfolio management (see
\cite{cover,helmbold}) the market is represented by a set of price
relative vectors $r_1,...,r_T \in \reals^n_+$. These vectors
represent the daily change in price for a set of $n$ assets. A
Constant Rebalanced Portfolio is an investment strategy that
redistributes the wealth daily according to a fixed distribution
$p \in \simplex$. A natural investment strategy computes the best
CRP up to a certain trading day and invests according to this
distribution in the upcoming day.

On this basic mathematical program many variants have been
proposed. In \cite{AH}, a logarithmic barrier function is added to
the objective, which enables to prove theoretical bounds on
the performance. Bertsimas \cite{bertsimas} suggested to add a
quadratic term to the objective function so to take into account
transaction costs. An example of a convex program to find the best
CRP, subject to transaction costs constraints is
\begin{align}
& \max \sum_{t=1}^T \log (p^\top r_t) + \sum_{i=1}^n \log (p^\top e_i) \label{eqn:portfolio2}  \\
& \|p - \tilde{p}\|_2^2 \leq c \notag \\
& p  \in \simplex \notag
\end{align}
The objective function includes the logarithmic barrier of
\cite{AH}, the vectors $\{e_i\}$ are the standard basis unit
vectors. The constraint enforces small distance to the current
distribution $\tilde{p}$ to ensure low transaction costs.

The Hessian of the objective is
$$ \sum_{t=1}^T \frac{1}{(p^\top r_t)^2} r_t r_t^\top + \sum_{i=1}^n \frac{1}{p_i^2} e_i e_i^\top \succeq I $$
The Hessian of the constraint is the identity matrix. Hence the
constant $H$ for Theorem~\ref{thm:scp} is one.

\subsection{Maximum entropy distributions for with $\ell_2$ regularization}

The following mathematical program arises in problems concerning
frequency estimation from a given sample. Examples include
modelling of species distributions \cite{miro} and haplotype
frequency estimation \cite{HH}.
\begin{align}
& \min H(p)  \label{eqn:miro}  \\
& \|A_i (p - \tilde{p})\|_2^2 \leq c \ \ i \in [m] \notag \\
& p  \in \simplex \notag
\end{align}
Where $H:\reals^n \mapsto \reals$ is the negative of the entropy function,
defined by $H(p) = \sum_{i=1}^n p_i \log {p_i}$. The Hessian of $H$ is
$\nabla^2 H(p) = diag(\frac{1}{p})$, i.e. the diagonal matrix with entries $\{
\frac{1}{p_i} , i \in [n]\}$ on the diagonal. Hence $\nabla^2 H(p) \succeq I$.

The hessian of the $i$'th constraints is $A_i A_i^\top$. For applications with
$\min_i A_i A_i^\top \succeq c \cdot I$, the constant $H$ for
Theorem~\ref{thm:scp} is $H = \min\{1,c\}$.

\section{Proof of Corollary \ref{cor:scp}}

\begin{proof}[proof of Corollary \ref{cor:scp}]
Given mathematical program (\ref{eqn:general}), we consider the following
program
\begin{align}
f_j (x) + \delta \|x\|_2^2 - \delta & \leq 0  \quad \forall j \in [m] \label{eqn:generalreduced} \\
x & \in \simplex \notag
\end{align}
This mathematical program has strictly convex constraints, as
$$ \forall i \in [m] \ . \ \nabla^2 (f_i(x) + \delta \|x\|_2^2 - \delta) =
\nabla^2 f_i(x) + 2 \delta I \succeq 2 \delta I$$ Where the last inequality
follows from our assumption that all constraints in (\ref{eqn:general}) are
convex and hence have positive semi-definite Hessian. Hence, to apply Theorem
\ref{thm:scp} we can use $H = 2 \delta$. In addition, by the triangle
inequality the gradients of the constraints of (\ref{eqn:generalreduced})
satisfy
$$ \| \nabla(f_i(x) + \delta \|x\|_2^2 - \delta) \|_2 \leq \|\nabla f_i(x)\|_2 +
2 \delta \leq G + 2 \delta = O(G) $$ Where $G$ is the upper bound on the norm
of the gradients of the constraints of (\ref{eqn:general}). Therefore,
Theorem~\ref{thm:scp} implies that a $\varepsilon$-approximate solution to
(\ref{eqn:generalreduced}) can be computed in $\tilde{O}(\frac{G^2}{\delta
\varepsilon})$ iterations, each requiring a single gradient computation and
additional $\tilde{O}(n)$ time.

\medskip

Notice that if (\ref{eqn:general}) is feasible, i.e there exists
$x^* \in S_n$ such that $\min_{i \in [m]} f_i(x^*) \leq 0$, then so is
(\ref{eqn:generalreduced}) since the same $x^*$ satisfies $\min_{i \in [m]}
f_i(x^*) + \delta \|x\|_2^2 -\delta \leq \delta \|x\|_2^2 - \delta \leq 0$.

Given a $\varepsilon$-approximate solution to (\ref{eqn:generalreduced}),
denoted $y$, it satisfies
$$ \forall j \in [m] \ . \ f_j (y) + \delta \|y\|_2^2 - \delta \leq \varepsilon \ \Rightarrow \ f_j(y) \leq - \delta \|y\|_2^2 + \delta + \varepsilon \leq \delta + \varepsilon $$
Hence $y$ is also a $(\varepsilon+\delta)$-approximate solution to
(\ref{eqn:general}).

Choosing $\delta = \varepsilon$, we conclude that a $2 \varepsilon$-approximate
solution to (\ref{eqn:general}) can be computed in
$\tilde{O}(\frac{G^2}{\varepsilon^2})$ iterations.
\end{proof}

\end{document}